\documentclass[10pt]{iopart}
\usepackage[english]{babel}
\usepackage{dcolumn}
\usepackage{bm}
\usepackage[utf8]{inputenc}
\usepackage{graphicx}
\usepackage{fancyhdr}
\usepackage[active]{srcltx}
\usepackage{setspace}
\usepackage{float}
\usepackage{color}
\usepackage{todonotes}
\begin{document}

\title{Electrical tuning of helical edge states in topological multilayers}

\author{T. Campos$^{1,4}$, M. A. Toloza Sandoval$^{2}$, L. Diago-Cisneros$^{3,4}$, G. M. Sipahi$^{4}$}
\address{$^1$ Department of Physics, State University of New York at Buffalo,\\ Buffalo, New York 14260, USA}
\address{$^2$ Instituto de F\'isica, Universidade Federal da Bahia,\\Salvador, Bahia 40210-340, Brazil.}
\address{$^3$ Facultad de F\'isica, Universidad de La Habana,\\ La Habana 10400, Cuba}
\address{$^4$ Instituto de F\'isica de S\~ao Carlos, Universidade de S\~ao Paulo,\\ S\~ao Carlos, S\~ao Paulo 13566-590, Brazil}

\begin{abstract}
Mainstream among topological insulators, GaSb/InAs quantum wells present
 a broken gap alignment for the energy bands which supports the quantum spin Hall
 insulator phase and forms an important building block in the search of exotic
 states of matter. Such structures allow the band-gap inversion with electrons and
 holes confined in adjacent layers, providing a fertile ground to tune the corresponding topological
 properties.
 Using a full 3D 8-band ${\bf k}\cdot{\bf p}$ method we investigate the inverted
 band structure of GaSb/InAs/GaSb and InAs/GaSb/InAs multilayers and the behavior of the helical edge states,
 under the influence of an electric field applied along the growth direction.
 By tuning the electric field modulus, we induce the change of the energy levels
 of both conduction and valence bands, resulting in a quantum spin Hall insulator phase where the helical edge states are predominantly confined in the GaSb  layer.
 In particular, we found that InAs/GaSb/InAs has a large hybridization gap of about
 $12\,\textrm{meV}$ and, therefore, are promising to observe massless Dirac
 fermions with a large Fermi velocity.
 Our comprehensive characterization of GaSb/InAs multilayers creates a basis platform
 upon which further optimization of III-V heterostructures can be contrasted.

\end{abstract}


\maketitle

\ioptwocol


\section{Introduction}

Conducting edge states are key pieces to understand the conceptual
puzzle behind the topological phases of
matter~\cite{qi2011x,Hasan2010}. Remarkably, unlike the integer
quantum Hall phase, which can be basically understood in terms of
chiral edge modes, quantum spin Hall insulators host time-reversal
protected helical edge states connecting valence and conduction
bulk bands -- the former case is fundamentally different from the latter
due to the presence of an external magnetic field that breaks the
time-reversal symmetry. Nevertheless, the spin Hall insulator phase was experimentally
observed, in InAs/GaSb and HgTe/(Hg,Cd)Te based heterostructures,
even for high magnetic
fields~\cite{du2015robust,ma2015unexpected}, and such unexpected
robustness of the gapless (edge) states forms a recent topic of
discussions~\cite{2017arXiv170904830S,li2018hidden,Krishtopenko2018}.

Over decades InAs/GaSb based structures have attracted much
attention in view of their unique properties, since the seminal
works exploring their unconventional band
alignment~\cite{Sakaki,sai1978inas}, where the top of the GaSb
valence band is higher than the bottom of the InAs conduction
band, forming a broken gap at the interface. In especial, such
broken gap alignment allows the collapse of the energy gap with
electrons and holes confined in adjacent layers, providing a
fertile testing-ground for fundamental and applied condensed
matter
physics~\cite{Ting1990,Chow,lakrimi1997minigaps,yang1997evidence,nicholas1998minigaps,Ting2002,kroemer20046}.
After a striking prediction~\cite{liu2008quantum}, the quantum
spin Hall insulator phase was experimentally observed in InAs/GaSb
heterostructures by different
groups~\cite{du2015robust,knez2011evidence,Knez2012,suzuki2013edge,nichele2014insulating,spanton2014images},
inspiring a race in the search of different exotic phases of
matter~\cite{Krishtopenko2018,marlow1999ground,naveh1996excitonic,pikulin2014interplay,pribiag2015edge,Li2015,Du2017,du2017evidence}.

Nowadays, there is a plethora of reports characterizing the edge
conductance behavior of InAs/GaSb asymmetric quantum well
(AQW)~\cite{du2015robust,knez2011evidence,Knez2012,suzuki2013edge,nichele2014insulating,spanton2014images,pribiag2015edge,Li2015,Du2017,
mueller2015nonlocal,karalic2016experimental,nichele2016edge,nguyen2016decoupling,Akiho2016,kononov2017proximity,mueller2017edge}.
Furthermore, electrical control of the topological phase transition is
possible~\cite{kim2012topological,liu2015switching,pan2015electric}
and has been reported for InAs/GaSb
AQWs~\cite{li2018hidden,qu2015electric,suzuki2015gate,hu2016electric,qi2016electrically}.
Although very recently Krishtopenko and
Teppe~\cite{Krishtopenko2018} proposed that three-layer InAs/GaSb
quantum well (QWs) hosts topological phase transition, their focus
was on the strain engineering of the band gap of the InAs/GaInSb
QWs which they claim is of the order of 60 meV. Instead, we focus on the electrical control, of the phase
transition and the behavior of the helical edge states in such
three-layer QWs, using an applied electric field but neglecting \textit{e-e}
interaction and disorder effects.

In one hand, the helical edges states with hidden Dirac point constitute a possible explanation
for the aforementioned robustness, on the other, they are described as massless Dirac 
fermions with an exactly linear dispersion only in the vicinity of the Dirac point (or small k). Additionally, we also show that it is possible to tune the inverted-band structure aiming
to obtain massless Dirac fermions with a large fermion velocity. In order to make this 
evaluation, we use a full 3D 8-band ${\bf k}\cdot{\bf p}$
method~\cite{Luttinger1955,Kane1966,enderlein1997fundamentals,winkler2003spin,Campos2018}
and employ the envelope function approximation to take into account the quantum confinement
~\cite{enderlein1997fundamentals,burt1999fundamentals}, together
with the plane wave expansion~\cite{ehrhardt2014multi,sipahi1996band}. In this way,
we were capable to have a full three dimensional solution of the system which is 
specially important to analyze, for example, the spatial distribution of the helical 
edge states along the confinement profile.

In the inverted band regime the hybridization gap, $\Delta E_h$, is opened at a finite wave vector, $k_c$,
with Fermi velocity $v_F \approx \frac{\Delta E_h}{2\hbar k_c}$~\cite{Du2017,hu2016electric}.
By changing the layer size and/or applying an external electric field, the value of $k_c$ and $\Delta E_h$
can be tuned. Usually, low values of $k_c$ (higher of $v_F$) are desired such that the bulk states
do not coexist with the edge states. Hence, a system where $k_c$ is as close as possible to the
$\Gamma$-point and the $\Delta E_h$ is as large as possible, minimizes all the rich
yet undesired physical phenomena that interferes with the edge states~\cite{marlow1999ground,naveh1996excitonic,pikulin2014interplay,Li2015,du2017evidence}. The estimation of the wave vector values in which the hybridization occurs gives the range $k_{c} \in [0.1, 0.3]\,\textrm{nm}^{-1}$ suggesting that the systems are in the deeply
inverted regime~\cite{Du2017}. Moreover, both InAs/GaSb AQWs and GaSb/InAs/GaSb symmetric quantum wells (SQWs)
have similar hybridization gaps, $\Delta E_h \approx 5\,\textrm{meV}$, leading to similar low values
of Fermi velocity $v_F \in [1, 4] \times 10^4\,\textrm{ms}^{-1}$ while InAs/GaSb/InAs SQWs have
$v_F \in [3, 9]  \times 10^4\,\textrm{ms}^{-1}$, $\Delta E_h \approx 12\,\textrm{meV}$ suggesting
that InAs/GaSb/InAs SQWs should be a better candidate to host massless Dirac fermions.

In summary, the race to build reliable platforms where the properties of the topological
phase can be efficiently harnessed is still on. One among several possible applications is to
build devices where Majorana fermions could be easily braided~\cite{Fatin2016,MATOSABIAGUE20171}.
Moreover, very recently it was shown that the hybridization gap of InAs/GaSb/InAs
multilayers is temperature independent~\cite{Krishtopenko2018temp} and also a clear evidence
of the massless Dirac fermions was reported~\cite{Krishtopenko2019}. Our main motivation and 
purpose is to demonstrate the possibility to tune multilayers as those depicted in \fref{figConf}
into a topological regime by applying an electric field, and this way enabling topological-edge 
states to arise. The theoretical modeling we present highlights the relevance of InAs/GaSb three-layer heterostructure for searching new topological phase transitions, due to the intrinsic alignment 
of quasi-bound electron states to quasi-bound hole levels as needed. We hope that our study 
could be used as a guidance in order to develop novel devices that use
switchable topological phase transitions.

\section{Topological Broken-Gap Multilayers}
\label{sec:top}

\begin{figure}[h!]
\includegraphics[width=0.5\textwidth]{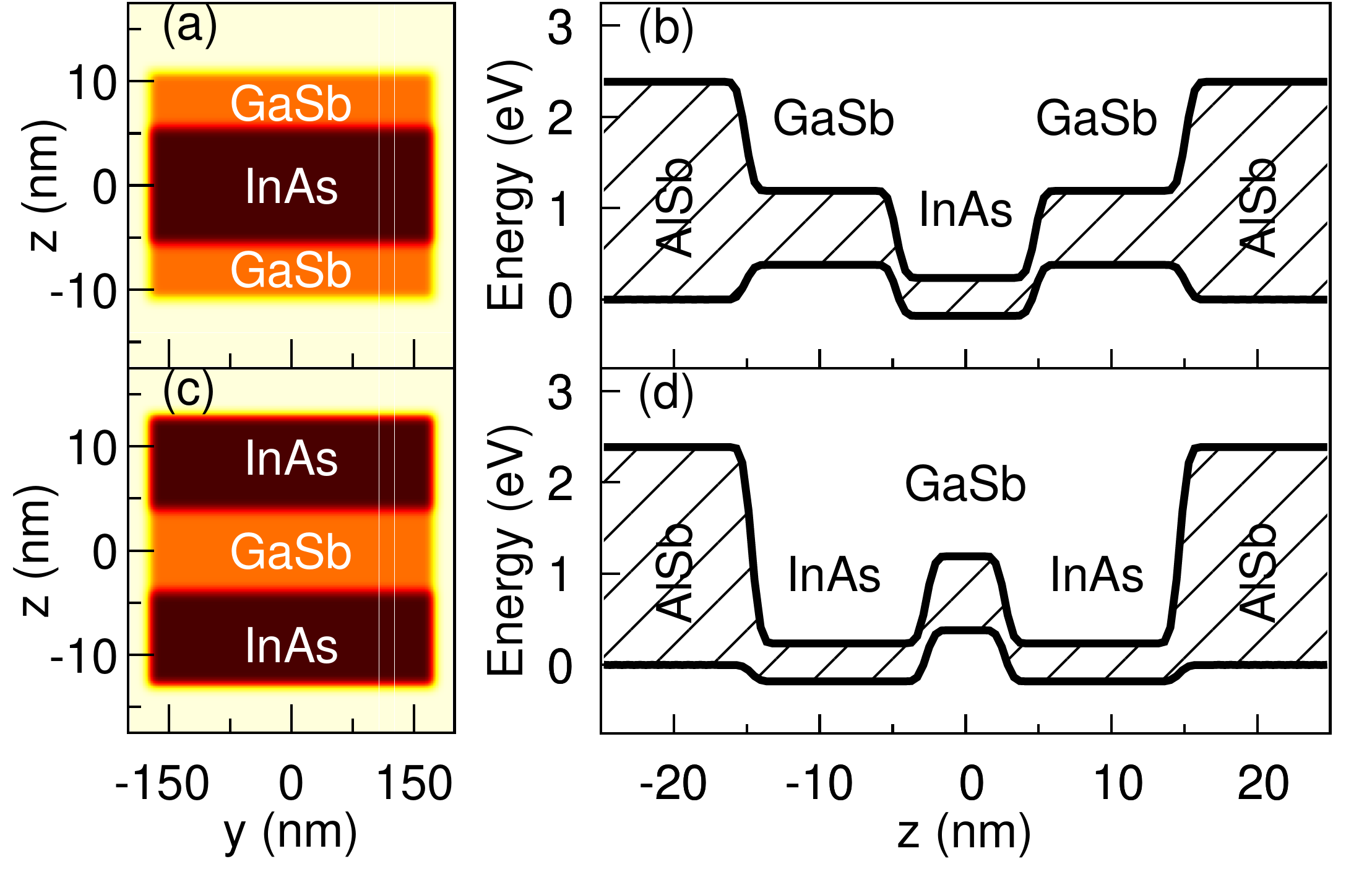}
\caption{Confinement profile of the broken-gap multilayers. (a)
and (b) slab configuration
 with quantum confinement along both $y$ and $z$ directions and quantum well
 confinement profile for the InAs/GaSb/InAs multilayer system; (c) and (d) slab configuration and quantum well
 confinement profile for the GaSb/InAs/GaSb multilayer system.}
\label{figConf}
\end{figure}

In this manuscript we study two distinct arrangements of broken-gap multilayers
that present topological features. The first system, GaSb/InAs/GaSb, consists of
one InAs layer surrounded by two GaSb layers, see \fref{figConf}(a)-(b), and
does not present a hybridization gap. The application of an external electric field
induces a Rashba spin-orbit coupling (SOC) that yields a hybridization gap
that, as we will show, have values close to the ones of a regular AQW of about $5\,\textrm{meV}$.
In the second system, InAs/GaSb/InAs, consisting of one GaSb layer surrounded
by two InAs layers, see \fref{figConf}(c)-(d), there is an anticrossing of
the electron and heavy-hole bands but no overall gap (see Supplementary Material).
When an electric field is applied the system now shows a hybridization gap of
the order of $12\,\textrm{meV}$ that is stable under changes of the applied electric field.

Although the existence of the hybridization gap is important for these systems, its
is only effective when this gap results in an overall gap, \textit{i. e.}, that stands for
all k points, appearing as a general feature in the DOS of the system.
For both configurations we analyze how the hybridization gap changes when Rashba
SOC is controlled by the external electric field, directly tuning the values of
the overall density of states (DOS) gap. We also analyze the quantum spin Hall
phase when we turn the system from a quantum well into a slab, by adding a weak
quantum confinement along one of the previous free directions. In \fref{figConf}
we depicted the confinement profile of both cases, quantum well and slab configuration.

The energies of the states can be tuned by changing the layer sizes. By changing them it
may be possible to align quasi-bound electron states of InAs to quasi-bound hole levels
of the GaSb, leading to conditions for resonance tunneling or even giant conductance
regime~\cite{ARIASLASO20121730}. Moreover, as the effective masses of electrons are
much smaller than the holes' counterparts, this tuning may be done more effectively
by changing the sizes of the InAs layer ($L_{\textrm{InAs}}$). Carefully adjusting
the layers sizes of the GaSb layers ($L_{\textrm{GaSb}}$) the hole's energy can cross
the electron's one and the system may be driven from an insulating
phase to a band inverted phase~\cite{Krishtopenko2018,Tsay1997,Sandoval_2017}.
In the rest of this article we explore the regimes in which such energy level crossings
leads to a topological phase transition by tuning the applied electric field.

\subsection{GaSb/InAs/GaSb multilayer}
\label{subsec:GaSb}

\begin{figure}[h!]
\includegraphics[width=0.5\textwidth]{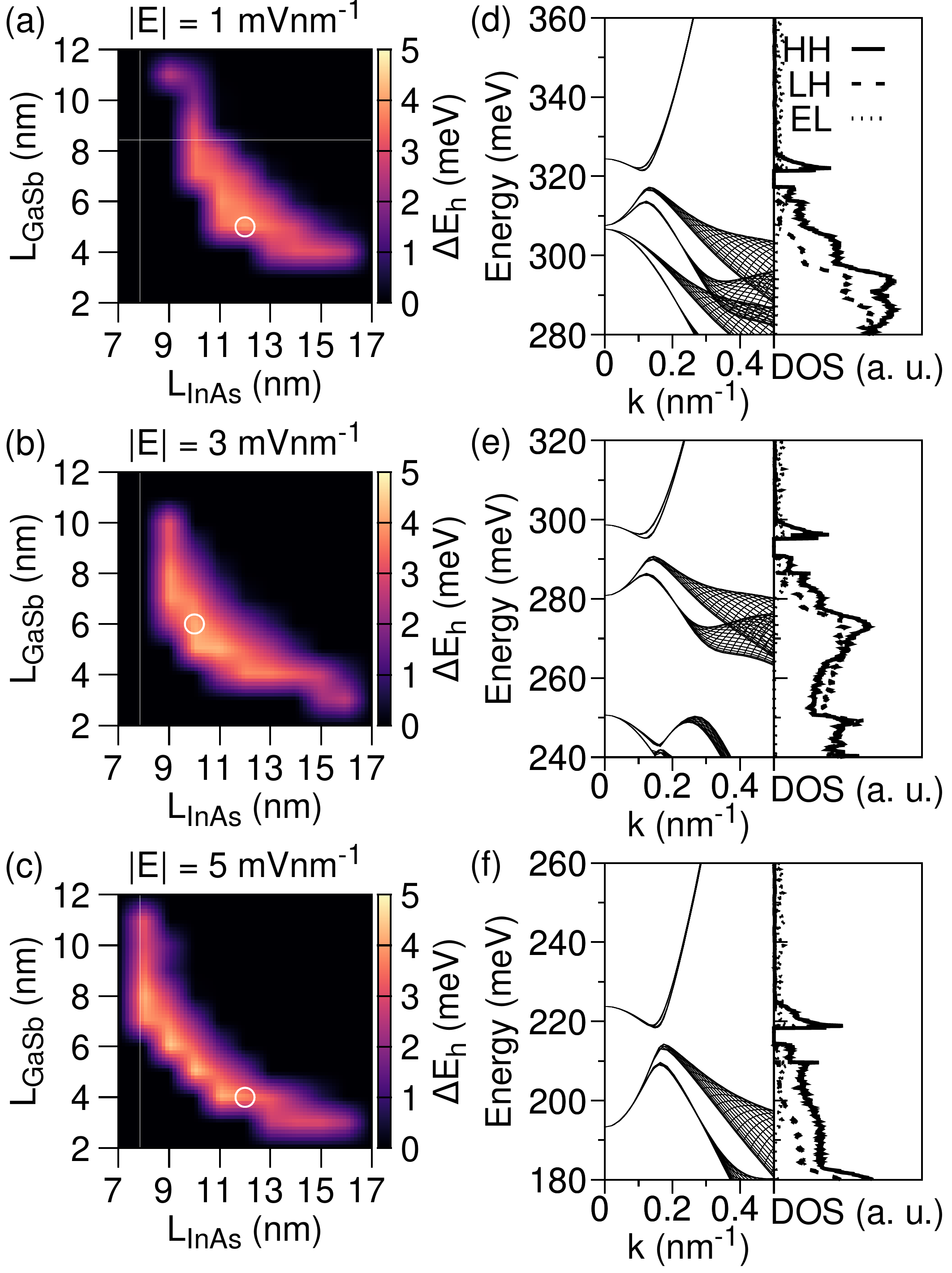}
\caption{Hybridization gap heatmap of the GaSb/InAs/GaSb SQW as a function of the InAs and GaSb
layer sizes. Through (a) to (c) we show the hybridization gaps for several values of external
electric field, the open circles indicate the sizes used to plot, through (d) to (f), 
the band structures and the projected DOSs after the band inversion. The label HH refers to heavy-hole, LH to light-hole and EL to electrons (or conduction band). More specifically, in (a)
$|E|=1\,\textrm{mV}\,\textrm{nm}^{-1}$, in (b) $|E|=3\,\textrm{mV}\,\textrm{nm}^{-1}$ and in
(c) $|E|=5\,\textrm{mV}\,\textrm{nm}^{-1}$. In (d)  $L_{\textrm{GaSb}}=5\,\textrm{nm}$ and
$L_{\textrm{InAs}}=12\,\textrm{nm}$, in (e) $L_{\textrm{GaSb}}=6\,\textrm{nm}$ and
$L_{\textrm{InAs}}=10\,\textrm{nm}$ and in (f) $L_{\textrm{GaSb}}=4\,\textrm{nm}$ and
$L_{\textrm{InAs}}=12\,\textrm{nm}$. The wave vector $k$ in (d), (e) and (f) is defined
as $k=\sqrt{k_x^2 + k_y^2}$.}
\label{fig01}
\end{figure}

The confinement profile of the GaSb/InAs/GaSb system leads to a spatial separation
of the carriers as seen in \fref{figConf}(b), electrons being confined in the
InAs layer and holes in the GaSb layers. Without breaking the system inversion
symmetry, the GaSb/InAs/GaSb SQW will only have a semiconductor-semimetal transition,
as discussed in the Supplementary Material.

Heatmaps showing the overall gap as a function of the $L_{\textrm{InAs}}$ and $L_{\textrm{GaSb}}$
are shown in  \fref{fig01}(a)-(c) for three different absolute values of the
electric field $|E|=1\,\textrm{mV}\,\textrm{nm}^{-1}$, $|E|=3\,\textrm{mV}\,\textrm{nm}^{-1}$ and
$E=|5|\,\textrm{mV}\,\textrm{nm}^{-1}$. Two different trends may be extracted from
these graphics. First, if the size of one layer and of the electric field are fixed, 
the variation of the size of the other layer induces a threshold in which the hybridization gap
opens. Further increasing the size of this layer, the gap reaches a maximum and decays
up to a second threshold where the gap closes, forming a crescent moon shape. Second,
if the applied electric field value is increased, the initial thresholds are moved
to smaller values of $L_{\textrm{InAs}}$ and $L_{\textrm{GaSb}}$ and the gaped region 
is compressed when $L_{\textrm{InAs}} \approx L_{\textrm{GaSb}}$ and stretched when 
they are more separated.

Both behaviors can be explained with a simple argument. The electric field ramp
changes the band edges differently for each layer and, consequently, also changes the relative
energy difference between states on different layers. For positive (negative) applied
electric fields, the outer holes states of the left (right) GaSb layer increases in
energy while the electron energy level decreases. A trivial band alignment, with electron
states having higher energies than holes, may become an inverted alignment for a sufficient
large electric field. Further increasing the electric field value may turn this
inverted regime system into a deeply inverted regime as in the InAs/GaSb
AQW~\cite{qu2015electric,hu2016electric}.

The deeply inverted regime can be engineered either by finding the right combination
of the layer sizes or by an applied electric field. From \fref{fig01}(a)-(c), we can
identify this regime by choosing a system configuration which is on the verge of
becoming gapless. One could also identify it by looking at the wave vector value
where the subbands anticross each other, $k_c$. For small magnitudes of the electric
field, the hybridization occurs closer to $\Gamma$-point and, therefore, the value
of $k_c$ is also smaller, compared to large values of electric field in the same system.
In our data, the values vary in the ranges of $k_{c} \in [0.1, 0.3]\,\textrm{nm}^{-1}$
and $v_F \in [1, 4] \times 10^{4}\,\textrm{ms}^{-1}$ and though, GaSb/InAs/GaSb SQW
shows a deep hybridization.

\Fref{fig01}(d)-(f) present the band structures and the projected DOSs for selected
deeply inverted regime SQWs, marked as open circles in \fref{fig01}(a)-(c), respectively.
The band structures were plotted as functions of all $k_{\parallel}$ setting $k=\sqrt{k_x^2 + k_y^2}$
allowing the visualization of features from all directions of the band structures
at once. The asymmetries of the valence band become very clear for wave vectors away
from $\Gamma$-point, showed as the spreading in energy of a given subband. In the
inverted regime the largest contribution in the projected DOS near the hybridization
gap region is due to the heavy-hole subband. Although the conduction band state should
be a linear combination of electrons and light-holes, as explicitly shown in simplified
models of topologically protected systems, like the Bernevig-Hughes-Zhang (BHZ) model~\cite{Bernevig2006}, the
light-hole contribution on the valence band is only relevant for wave vectors away from the $\Gamma$-point
and/or energies well below the hybridization region. 
This means that in order to
correctly describe the low-energy spectrum of such SQW there is no need to include
the light-hole band, as it is important for the InAs/GaSb AQW~\cite{hu2016electric,li2018hidden},
although we need to include the extra heavy-hole band~\cite{Krishtopenko2018}.

The effect of the electric field in expelling the lower heavy-hole subbands can
be seen in \fref{fig01}(d), \ref{fig01}(e) and \ref{fig01}(f). In \fref{fig01}(d)
both heavy-hole subbands are almost degenerate at $\Gamma$ with energies close to 310 meV.
The spin-splitting of each subband is seen at higher $k_\parallel$. In fig \ref{fig01}(e),
the second heavy-hole appears at around 250 meV and in \fref{fig01}(f) it is out of the range.
Therefore, in the limit of very large electric fields the system becomes similar to the
InAs/GaSb AQW, \textit{i. e.}, the electric field isolates one of the GaSb layers from
the GaSb/InAs and simpler models such as the aforementioned BHZ model~\cite{Bernevig2006}
can describe very accurately the low energy properties of the system.

\begin{figure*}[ht]
    \includegraphics[width=\textwidth]{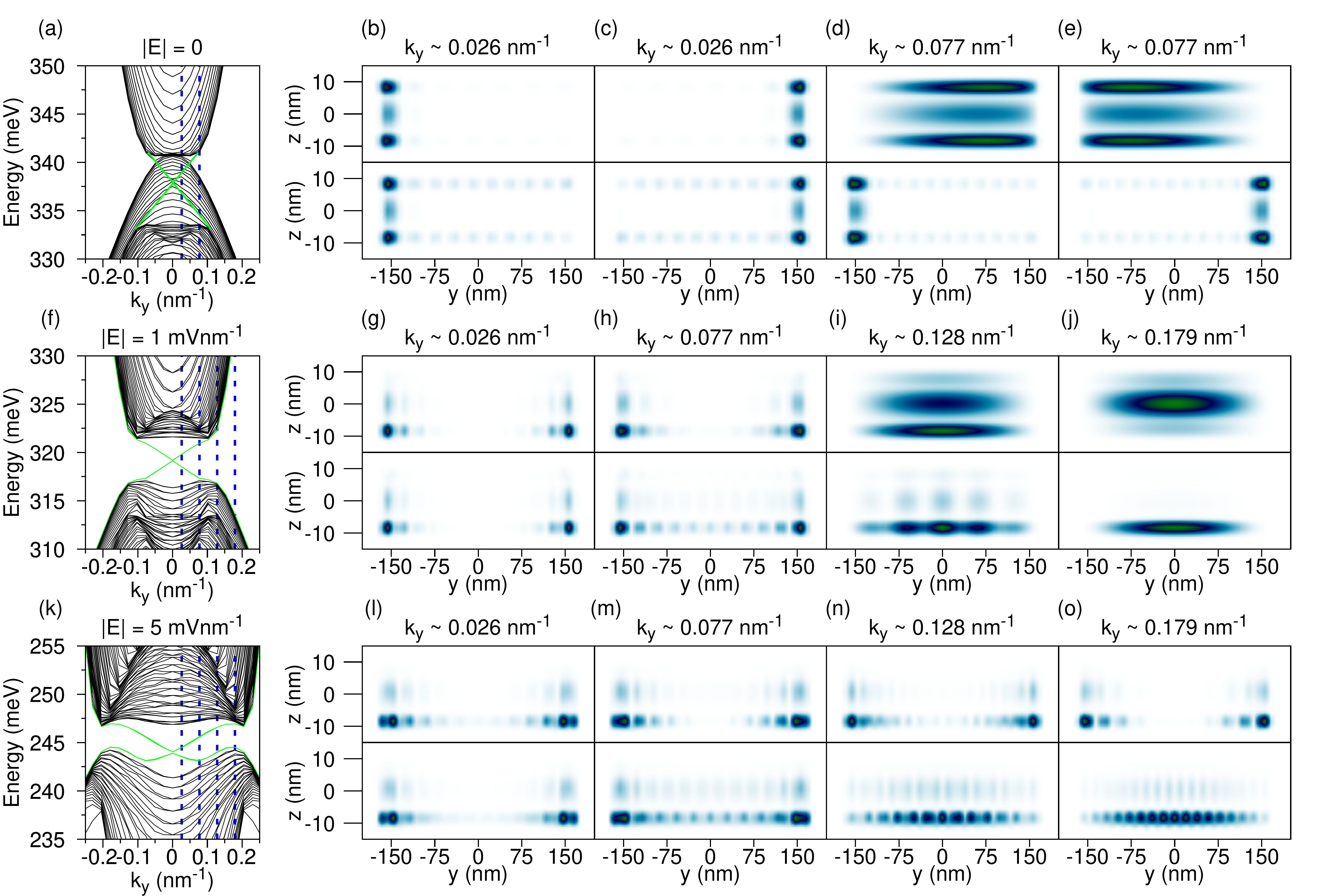}
    \caption{Energy dispersion and edge states probability densities for a
        $L_{\textrm{GaSb}}=5\,\textrm{nm}$ and $L_{\textrm{InAs}}=12\,\textrm{nm}$ SQW.
        (a)-(e) are for $E = 0$. (f)-(j) are for $|E|=1\,\textrm{mV}\,\textrm{nm}^{-1}$.
        (k)-(o) are for $|E|=5\,\textrm{mV}\,\textrm{nm}^{-1}$. 
        The green lines in (a), (f) and (k) are guidelines 
        to highlight where the Dirac cone like dispersion is. The blue lines in the same panels show the wave vectors depicted in the right hand panels.}
    \label{fig02}
\end{figure*}

\subsubsection*{Massless Dirac fermions\\}
\label{subsec:GaSbDirac}

The investigation of the Dirac cones like energy dispersion and edge states is 
usually done by using effective models that include a limited number of bands, 
such as the 4-band BHZ model~\cite{Bernevig2006} and other extended BHZ-like models 
that include extra subbands~\cite{li2018hidden,Krishtopenko2018}. The advantage of 
these models is the number of states included, requiring smaller computational resources 
to be numerically solved. What makes them simpler, in few words, is that symmetry is
used to define a basis, with a smaller number of states that describes the essential
features of the system in the confinement direction.
This basis, composed by states that are linear combinations of components of the
original basis specifically tuned to describe the intended few subbands near the gap, 
is used to generate an effective Hamiltonian that has the confinement direction 
integrated out and replaced by a set of parameters.
The integration that allows to simplify the description is also its major drawback,
since the spatial resolution of the probability densities along the quantum
well confinement is lost. Here, instead of using the BHZ-like models, we use the full
3D 8-band ${\bf k}\cdot{\bf p}$ model, and therefore, we keep the spatial resolution
along the quantum well confinement axis.

Analyzing the probability densities one can see that as we increase the momentum, $k_y$, they
transition from a edge localized state to a sample centered localized state. Another
feature is that the edge states are oscillating~\cite{li2018hidden}, \textit{i. e.}, 
they have peaks and nodes along the $y$ direction, but in general the density presents
a very intense peak near the edge and the intensity of the peaks toward the center diminish, forming a tail. In the Supplementary Material we briefly go over the reason why such oscillations are present, a more detailed explanation is given in Ref. \cite{Erlingsson2015}.

In \fref{fig02}(a) the system has a linear energy dispersion band (emphasized by the green 
line) but it is embedded on the valence subbands, \textit{i. e.}, there is an edge state that connects
the conduction band to the second valence band~\cite{Krishtopenko2018}. Although such edge 
state exists, the system is in a semimetal phase since no overall gap is present. To verify
that this embedded linear energy dispersion corresponds to edge states, we plot in
\fref{fig02}(b)-(e) the probability density of such states, for the conduction band (upper panels) and valence band (lower panels) at $k_y = 0.026\,\textrm{nm}^{-1}$ ((b) and (c))
and at $k_y = 0.077\,\textrm{nm}^{-1}$ ((d) and (e)). Notice first that each state
is doubly degenerate in spin. Therefore, in \fref{fig02}(b) and \ref{fig02}(c)
we plot the probability density of one of the spin components, while in \fref{fig02}(d)
and \ref{fig02}(e) we plot the other spin component, for a given wave vector. It
is evidently clear that these states are helical edge states since each spin projection
is confined along an opposite edge of the slab. Moreover, by using the full 3D 
${\bf k}\cdot{\bf p}$ description we are able to see that the edge states are symmetric
with respect to the InAs layer (central layer), since there is no electric field 
breaking this symmetry, and are predominantly located into the GaSb layers. For
$k_y = 0.077\,\textrm{nm}^{-1}$ the conduction band edge state is still predominantly
in the GaSb layer but it is also more spread throughout the slab and does not present
oscillations indicating that it is losing its edge state character and is becoming a bulk-like state.
The existence of a Dirac cone inside the valence band continuum may be understood as Bound State 
in the Continuum, or BIC, phenomena~\cite{hsu2016bound}. In this specific case a symmetry protect
BIC emerges inside the continuum of the valence band due to the presence of a reflection
symmetry.

In \fref{fig02}(f)-(j) we show the energy dispersion and the probability densities for the
same GaSb/InAs/GaSb multilayer with applied electric field of
$|E|=1\,\textrm{mV}\,\textrm{nm}^{-1}$. In \fref{fig02}(f) we see that the second 
heavy-hole band is no longer on the same energy range as the Dirac cone like energy 
dispersion and that it has become very visible and well isolated from the other subbands.
The probability  densities are shown on \fref{fig02}(g)-(j), but now at four distinct
wave vector values and its worth noticing that, although we are only showing one of
the spin components, they have peaks at both interfaces with opposite spin projections
(as it should be for helical edge states)~\cite{Budagosky2017}. Since there is an
applied electric field, the states are not anymore symmetric with respect to the
InAs layer, however they still are predominantly confined on the GaSb layers. The
probability densities at $k_y = 0.026\,\textrm{nm}^{-1}$ and $k_y = 0.077\,\textrm{nm}^{-1}$
show the edge states with oscillation and light tails as seen in \fref{fig02}(g)
and \ref{fig02}(h). As the wave vector increases however, the overall behavior of
these oscillations change and the more intense peaks start spreading from the edges
to the center of the slabs, as seen in the valence band panel of \fref{fig02}(h).
Further increasing the wave vector, the oscillations end on the conduction band, even
though the state still has a higher confinement on the GaSb layer -- but persists on
the valence band but now the oscillations are in small number and more intense at the
center of the confining potential, as seen in \fref{fig02}(i). This mixed state is on the
transition from an edge state to a bulk state and we can see that its wave vector
is at the point where the Dirac cone dispersion touches the bulk-like energy dispersion. 
Choosing the wave vector away from the Dirac cone like energy dispersion region, 
see \fref{fig02}(j), the states become bulk-like states in which the conduction band has
its peak on the InAs layer and the valence band on the GaSb layer.

A striking feature of some quantum spin Hall systems, such as InAs/GaSb AQW and
HgTe/(Hg,Cd)Te QWs, is that it was experimentally observed that they host a robust
helical edge state which persists up to an applied magnetic field of several
Tesla~\cite{du2015robust,ma2015unexpected}. This robustness was latter on explained
by the burying of the Dirac cone energy dispersion into the valence
subbands~\cite{li2018hidden,2017arXiv170904830S}.
Here, we also show that proposed SQW quantum spin Hall system also has such buried Dirac
cone, and therefore hosts a robust edge that should persists up to several Tesla.
In general, for such feature to be present in the SQW it must be in the deeply
inverted regime. Therefore, if a selected system does not present the buried Dirac cone 
dispersion, a recipe to achieve it is to increase the applied electric field, thus
enhancing the inverted regime. In \fref{fig02}(k) we can see that by increasing the applied
electric field we push the minimal energy difference between the subbands away from the
$\Gamma$-point (it is at 0.0 $\textrm{nm}^{-1}$ in (a), a little below 1.28
$\textrm{nm}^{-1}$ in (f) and close to 0.2 $\textrm{nm}^{-1}$ in (k)), meaning that (k)
is on a deeply inverted regime. Focusing on the right panel of \fref{fig02}(k), one can
see that, indeed, the Dirac cone dispersion was pushed down towards the bulk 
valence subbands and it is hidden.

In \fref{fig02}(l)-(o) we show the probability densities of four selected wave
vector values. The behavior as we increase the $k_y$ value is similar to that we discussed
early. The only distinction is that the valence band branch of the Dirac cone dispersion touches the bulk valence bands while inside the hybridization region. This changes the probability densities from a edge localized to bulk like but still with the oscillations, as we can see in \fref{fig02}(m)-(o), while the conduction band branch still shows edge localized states.

In summary, the hybridization gap for the GaSb/InAs/GaSb SQW have a similar magnitude as the
InAs/GaSb AQW. The presence of an extra heavy-hole subband, due to the second GaSb layer,
precludes the opening of the hybridization gap and by tuning the layer sizes, the system 
only presents a semiconductor-semimetal transition. The application of an external electric
field is therefore a requirement to open the hybridization gap. Moreover, estimations of
$k_c$ and $v_F$ indicate that this system is in a deeply inverted regime. By confining
along the $y$ direction and analyzing the spatial distribution of the edge states we
discovered that they are predominantly confined at the GaSb layer. As we could see, in
\fref{fig02}, as we increase the applied electric field and enhance the inverted regime,
the probability densities become more oscillating and its overall maximum go further to
the middle of the confining profile.

\subsection{InAs/GaSb/InAs multilayer}
\label{subsec:InAs}

\begin{figure}[h]
\includegraphics[width=0.5\textwidth]{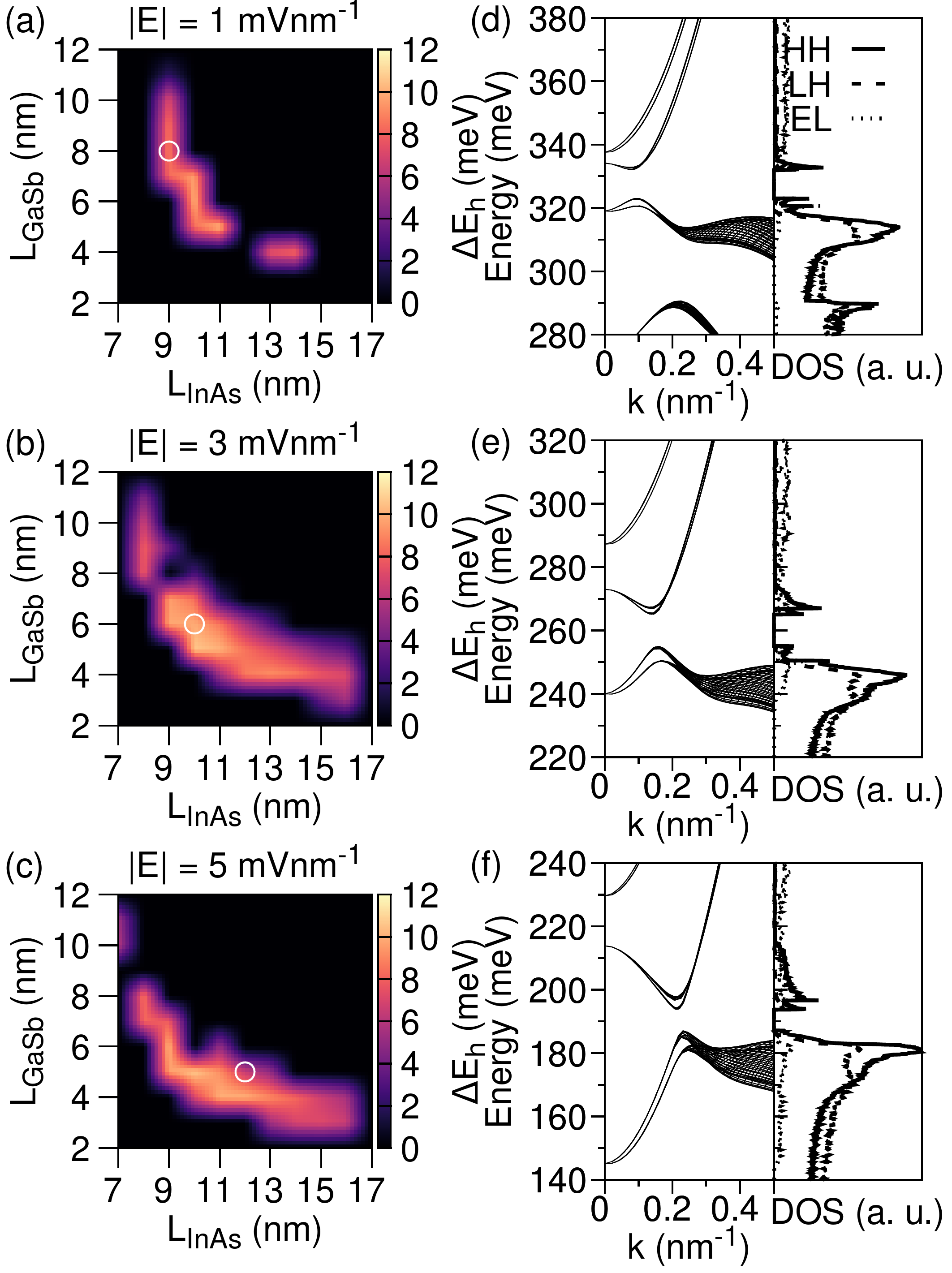}
\caption{ Same as \fref{fig01} but in (d) $L_{\textrm{GaSb}}=8\,\textrm{nm}$ and $L_{\textrm{InAs}}=9\,\textrm{nm}$, in (e) $L_{\textrm{GaSb}}=6\,\textrm{nm}$ and $L_{\textrm{InAs}}=10\,\textrm{nm}$ and in (f) $L_{\textrm{GaSb}}=5\,\textrm{nm}$ and $L_{\textrm{InAs}}=12\,\textrm{nm}$.}
\label{fig03}
\end{figure}

In the InAs/GaSb/InAs configuration, electrons are confined in the lateral wells
and the holes inside the central well as seen in \fref{figConf}(d). As shown in
the Supplementary Material, the semiconductor-semimetal transition occurs as a result
of the interaction of the subbands that gradually shifts the top of the valence band 
away from $\Gamma$-point while at the same time the electron subbands penetrates 
the valence ones. In this section we explore the phase diagram of the InAs/GaSb/InAs SQW.

\Fref{fig03}(a)-(c) present heatmaps with the complete phase diagram of the hybridization
as a function of $L_{\textrm{InAs}}$ and $L_{\textrm{GaSb}}$ for the same three absolute
values used in InAs/GaSb/InAs systems: $|E|=1\,\textrm{mV}\,\textrm{nm}^{-1}$,
$|E|=3\,\textrm{mV}\,\textrm{nm}^{-1}$ and $|E|=5\,\textrm{mV}\,\textrm{nm}^{-1}$.
Similarly to the GaSb/InAs/GaSb SQW, the hybridization gap occurs for distinct
values of the layers since the electric field counteracts the effect of the quantum
confinement by decreasing (increasing) the energy of the conduction (valence) states.
Also similarly to the previous SQW $k_c \in [0.1, 0.3]\,\textrm{nm}^{-1}$ and
$n_c \in [2, 14] \times 10^{11} \,\textrm{cm}^{-2}$. However, since $\Delta E_h$ is larger
the $v_F \in [3, 9] \times 10^4\,\textrm{ms}^{-1}$ meaning that the InAs/GaSb/InAs SQW is
not so much in a deep inverted regime compared to InAs/GaSb AQW and GaSb/InAs/GaSb SQW.

\Fref{fig03}(d)-(f) presents the band structures and the projected DOSs for selected
representative multilayers, marked as open circles in \fref{fig03}(a)-(c), respectively.
Analogous to the previous case, the asymmetries of the valence band become clear for wave
vectors away from $\Gamma$-point and are shown as the energy spreading of a given subband
and again, in the inverted regime, the largest contribution in the projected DOS
near the hybridization gap region is from the heavy-hole band. The light-hole contribution
is only relevant for wave vectors away from the $\Gamma$-point and/or energies well
below the hybridization region. This means that in order to correctly describe the
low-energy spectrum of such SQW there is no need to include the light-hole band
although the need to include one extra electron band~\cite{Krishtopenko2018}.

The effect of the electric field in expelling the electron subbands can be seen in
\fref{fig03}(d)-(f), in which for a small electric field, the second electron
subband is less than $10\,\textrm{meV}$ distant from the first and for
$|5|\,\textrm{mV}\,\textrm{nm}^{-1}$ is about $20\,\textrm{meV}$. In the limit of
very large electric fields, the system becomes similar to the InAs/GaSb AQW, \textit{i. e.},
the electric field isolates one of the InAs layers from the other GaSb and InAs layers.
Therefore, simpler models such as the BHZ model~\cite{Bernevig2006} can describe
very accurately the low energy properties of the system.

\begin{figure*}[ht]
    \includegraphics[width=1\textwidth]{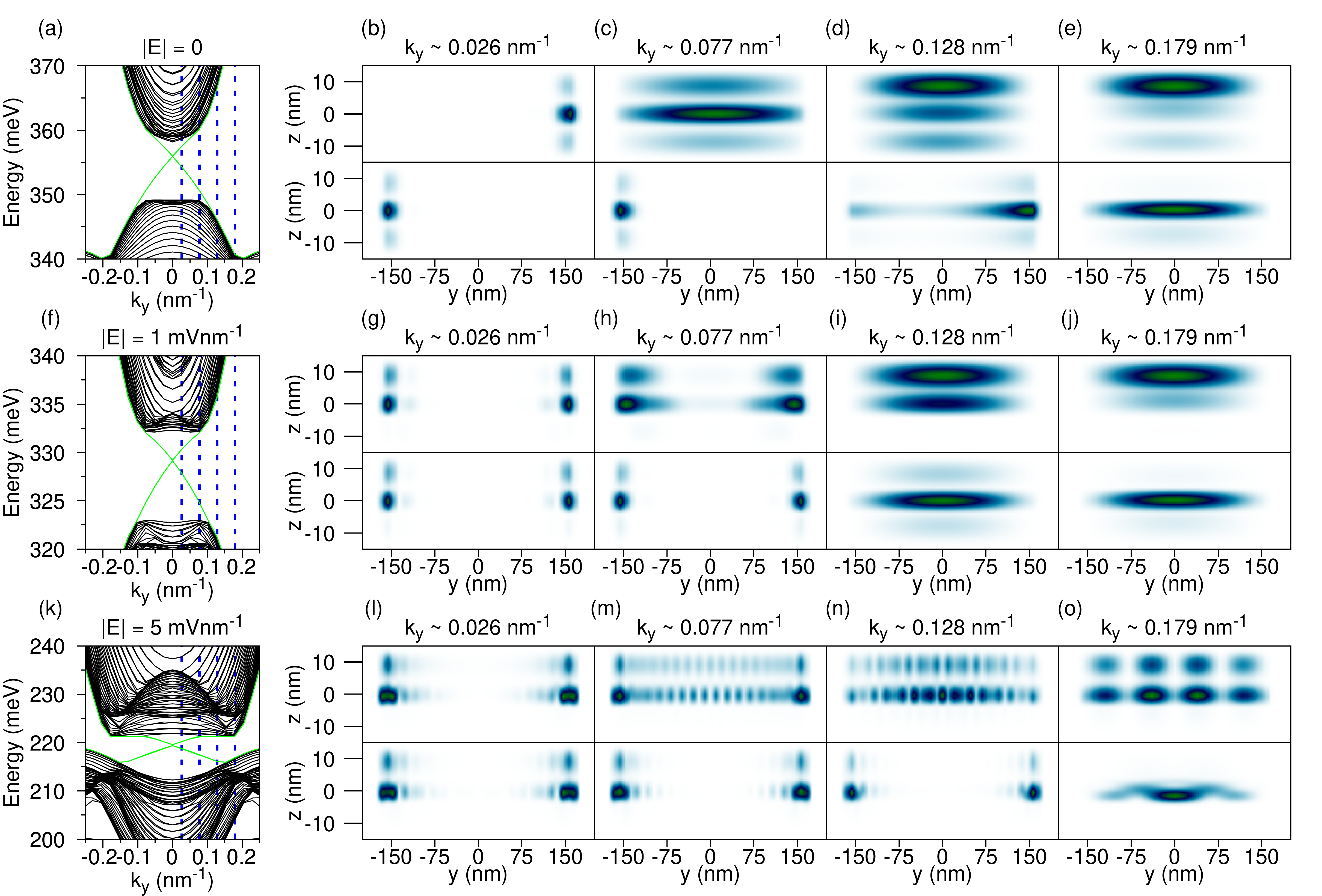}
    \caption{Energy dispersion and edge states probability densities for a
        $L_{\textrm{GaSb}}=8\,\textrm{nm}$ and $L_{\textrm{InAs}}=9\,\textrm{nm}$ SQW.
        (a)-(e) are for $E = 0$. (f)-(j) are for $|E|=1\,\textrm{mV}\,\textrm{nm}^{-1}$.
        (k)-(o) are for $|E|=5\,\textrm{mV}\,\textrm{nm}^{-1}$. 
        The green lines in (a), (f) and (k) are guidelines to highlight where the Dirac 
        cone is. The blue lines in the same panels show the wave vectors depicted in the 
        right hand panels.}
    \label{fig04}
\end{figure*}

\subsubsection*{Massless Dirac fermions\\}
\label{subsec:InAsDirac}

In \fref{fig04}(a) we show the band structure for the case without applied electric field.
We can see that due to the conduction and valence band hybridization, as discussed in the
Supplementary Material, the system shows a very isolated Dirac cone energy dispersion. \Fref{fig04}(b)-(e) shows the probability densities of the edge states at
four selected wave vector values. Notice that contrary to the previous case, the
linear dispersion here is not embedded on the valence subbands and instead it is
isolated at the hybridization gap energy region. Only one of the spin components
of each state is shown since the other component is its conjugate.

For $k_y=0.026 \textrm{nm}^{-1}$, the edge states are well
confined at the edges of the system and the oscillatory tail towards the middle of
the slab have almost zero intensity, as seen in \fref{fig04}(b). Moving away, for
$k_y=0.077 \textrm{nm}^{-1}$ (\fref{fig04} (c)), the conduction band states
have already become bulk-like states with the probability density maximum on the GaSb
layer while the valence states are yet confined at the edges. For 
$k_y=0.128 \textrm{nm}^{-1}$ (\fref{fig04}(d)) the states are outside the Dirac cone
dispersion region (see \fref{fig04}(a)). In this case, the conduction band state is
bulk-like and now mostly confined at the InAs layer while the valence band still
shows confinement near the edges but with non-zero probability density all over the slab. 
Finally for $k_y=0.179 \textrm{nm}^{-1}$, in \fref{fig04}(e), all the states are bulk 
like, with both conduction and valence states confined at the center of the confining 
potential.

The application of an external electric field accentuates the edge states features.
From the previous section, we know that it pushes the second electron subband away
from the hybridization energy range and the $k_c$ is pushed further away from the
$\Gamma$-point. \Fref{fig04} (g)-(j) show probability densities for selected
wave functions in a system with $|E| = 1\,\textrm{mV}\,\textrm{nm}^{-1}$. For
$k_y=0.026\,\textrm{nm}^{-1}$, the states are very confined at the edges with
almost no oscillatory tail towards the center of the slab, as seen in \fref{fig04}(g).
\Fref{fig04}(h) shows that both, conduction and valence band states at
$k_y=0.077\,\textrm{nm}^{-1}$, are still edge states. However, the conduction band
density probability has a non-oscillatory stronger tail towards the center of the
slab and both states are more localized in the GaSb region. Moving to 
$k_y=0.128\,\textrm{nm}^{-1}$ and $k_y=0.179\,\textrm{nm}^{-1}$, we get away from the 
Dirac cone dispersion region and, indeed, the states becomes bulk-like and conduction
band states become localized at one of the InAs regions, as seen in \fref{fig04}(i)
and \ref{fig04}(j).

The estimations of the Fermi velocities for the InAs/GaSb/InAs SQW,
$v_F \in [3, 9]  \times 10^4\,\textrm{ms}^{-1}$, indicates that its inverted regime
is not as deep as the GaSb/InAs/GaSb SQW. Indeed, by analyzing the probability densities
of both SQWs we gather that the former has well localized edge states while the 
latter has edge states with a highly oscillating tail. Moreover, the Fermi velocity is
dependent on the hybridization gap and a larger $\Delta E_h$ (which is the case of InAs/GaAs/
InAs) gives a larger velocity, therefore making this a better candidate to host 
experimentally detect the massless Dirac fermions.

By increasing the applied electric field we can tune the system to a deeply inverted
regime, as discussed before, where the Dirac cone becomes buried in the bulk valence
subbands. In \fref{fig04}(k) we present the energy dispersion showing a distinct
case since the top of the valence band is at a finite wave vector away from $\Gamma$-point,
almost engulfing the conduction band. In such a situation a hidden Dirac cone is
developed. Focusing on the right panel of \fref{fig04}(k) one indeed see that the
Dirac cone like energy dispersion is hidden.

\Fref{fig04}(l)-(o) show the probability densities of four selected wave vector values.
For $k_y=0.026\,\textrm{nm}^{-1}$ the edge states are very confined at the edges with a
slightly oscillating tail, as seen in \fref{fig04}(l). As the wave vector increases, the
oscillations of the conduction band edge state are accentuated, but the valence band 
edge state remains unchanged, as seen in \fref{fig04}(m). In \fref{fig04}(n), for
$k_y=0.128\,\textrm{nm}^{-1}$, the conduction band edge state has switched to a 
more bulk-like state with a higher probability to be found in the center but still
with an oscillating pattern, while the valence band is still well localized at the edge.
Moving to $k_y=0.179\,\textrm{nm}^{-1}$, see \ref{fig04}(o), we are still inside the
hybridization region, although both conduction and valence states had acquired a more bulk like shape they are not yet fully bulk states. 

In summary, the hybridization gap for the InAs/GaSb/InAs SQW is about $12\,\textrm{meV}$
and larger than both GaSb/InAs/GaSb SQW and InAs/GaSb AQW and the estimation of
$k_c$ and $v_F$ indicates that the system is not so deep inverted as its counterparts.
By confining along the $y$ direction and analyzing the spatial distribution of the
edge states it is shown that they are predominantly confined at the GaSb layer.
As we could see, in \fref{fig04}, the edge states are very localized at the edges
with very small oscillating tail, if any. By increasing the electric field we could
transition the system to a deeply inverted one with a hidden Dirac cone.

\section{Modeling details}
\label{sec:model}

To correctly account for the hybridization of the conduction and
valence bands an accurate description of the valence band states is
needed, specially for the heavy- and light-hole states. The usual 8-band Kane
model~\cite{Kane1966} is not adequate for this description since it gives the
wrong value for the effective mass of the heavy-hole states in
certain materials~\cite{bastard}. In this work we used the 8-band ${\bf k}\cdot{\bf p}$
model including the conduction band states, the three valence bands, the
explicit coupling between them and the Luttinger corrections for
the effective masses~\cite{Luttinger1955,enderlein1997fundamentals,winkler2003spin}.
This model has been successfully applied in describing the electronic
and spintronic properties of low dimensional semiconductor for decades~\cite{sipahi1996band,rodrigues2000valence,rodrigues2004charge,faria2012band,Cuan_2015,Campos2018}, including the InAs/GaSb AQWs.~\cite{li2018hidden,2017arXiv170904830S,halvorsen2000optical,zakharova2002strain,lapushkin2004self,
andlauer2009full,hong2009applicability,li2009spin,xu2010band,beukman2017spin,ndebeka2017disorder,jiang2017probing}

The ${\bf k}\cdot{\bf p}$ model describes a bulk material around a
high symmetry point, usually the $\Gamma$-point, where the physics of
interest takes place. To correctly describe it, a parametrization of
the matrix elements among the basis states (Bloch functions) is derived
using Group Theoretical methods. To describe a heterostructure, as a QW, we
apply the envelope function approximation~\cite{burt1999fundamentals,bastard},
that defines the total wave function of a state of the system as a continuous
and slowly varying function, called the envelope function, that is
weighted by the Bloch's function of each material. The quantum
confinement along the growth direction essentially means that we
have to make the substitution $k_z \rightarrow -i \frac{\partial}{\partial z}$.
This approach results in a system of coupled linear differential equations.
This system is solved by applying the plane wave expansion, using the Fourier
transformations
-- using 40 planes waves which suffices to achieve energy convergence in our calculations. The method is well described in our previous works~\cite{rodrigues2000valence,Campos2018}
We solve the final matrix Hamiltonian by direct diagonalization methods using the
MAGMA~\cite{dghklty14} suite which implements the LAPACK routines in a
multicore + GPU (graphical processing unit) computational environment.

The parameters used in the 8-band ${\bf k}\cdot{\bf p}$ model were
extracted from Ref. \cite{Vurgaftman}. It is know that solution of
narrow gap semiconductors is plagued with spurious solutions and to
avoid it we apply a renormalization of the Kane interband momentum matrix
element, $P$, as suggested by Ref. \cite{foreman1997elimination},
\begin{equation}
P^{2}=\left(\frac{m_{0}}{m_{c}}-1\right)\frac{E_{g}\left(E_{g}+\Delta\right)}{E_{g}+\frac{2}{3}\Delta}\frac{\hbar^{2}}{2\,m_{0}}
\end{equation}
where we set $\frac{m_{0}}{m_{c}} = 1$ with $E_g$ being the gap energy and $\Delta$
the spin-orbit splitting energy. With the new $P$ we then calculate
the new corrected Luttinger parameters according to

\begin{eqnarray}
\tilde{\gamma_{1}} & = & \gamma_{1}-\frac{E_{P}}{3E_{g}}\nonumber \\
\tilde{\gamma_{2}} & = & \gamma_{2}-\frac{E_{P}}{6E_{g}}\nonumber \\
\tilde{\gamma_{3}} & = & \gamma_{3}-\frac{E_{P}}{6E_{g}}\nonumber \\
A & =& \frac{1}{m_{e}^{*}}-\left(\frac{E_{g}+\frac{2}{3}\Delta_{\textrm{SO}}}{E_{g}+\Delta_{\textrm{SO}}}\right)\frac{E_{P}}{E_{g}}
\label{eq:LuttPar}
\end{eqnarray}
\\ where $E_{P} = \frac{2\,m_{0}}{\hbar{{}^2}}P^{2}$.

\section{Conclusion}
\label{sec:conclusion}

In this manuscript we have explored the electric control of the topological phase
transition in GaSb/InAs/GaSb and InAs/GaSb/InAs symmetric multilayers and the
spatial distribution of the edge states probability density using a full 3D
8-band ${\bf k}\cdot{\bf p}$ method. 

We have calculated the full hybridization gap phase diagram by varying the layer size 
and applied electric field. We have shown that the hybridization gap for the 
GaSb/InAs/GaSb SQW have values around $5\,\textrm{meV}$ that are very similar to the
InAs/GaSb AQW, while for InAs/GaSb/InAs SQW is about $12\,\textrm{meV}$. In both 
systems the hybridization occurs at similar values of wave vector 
$k_c \in [0.1, 0.3]\,\textrm{nm}^{-1}$, therefore the InAs/GaSb/InAs SQW having a 
large hybridization gap will also have a larger Fermi velocity. Ultimately, this 
means that the edge states of the InAs/GaSb/InAs SQW are less interacting with 
the bulk bands.

By applying a weak confinement along the $y$ direction and analyzing the spatial
distribution of the edge states, we verified that they are predominantly confined
at the GaSb layer. By increasing the electric field we tune the system to a deeply
inverted regime with a hidden/buried Dirac cone like energy dispersion having highly 
oscillating probability densities with large tails towards the bulk. Although this feature is 
present in both multilayers, the InAs/GaSb/InAs SQW due to its large hybridization gap,
suppress it, and therefore is suggested as the better option to explore the rich physics
offered by the quantum spin Hall.

Departing from the two systems under consideration, we have unambiguously demonstrated 
the relevance of the InAs/GaSb three-layer heterostructures. Being a platform extremely
tunable, they can be used as a playground to further expand out scientific knowledge 
of topics ranging from many-body interactions~\cite{pikulin2014interplay,Li2015,du2017evidence}
-- tuning $\Delta E_h$ to very small values -- to quantum computing with Majorana
fermions~\cite{Fatin2016,MATOSABIAGUE20171} -- tuning $\Delta E_h$ to large values.

\ack
This work has been partially supported by CAPES (CsF - grant No. 88881.068174/2014-01
and PNPD - grant No. 88882.306206/2018-01) and PLAF / SBF / CNPq. The first author
thanks the LCCA for computational resources, D. R. Candido and P. E. Faria Junior
for useful insights and discussion. L. D-C gratefully acknowledge the hospitality
of IFSC-USP.

\section*{References}
\bibliography{./ref}

\end{document}